\documentclass[aps,prl,showpacs,reprint,longbibliography,floatfix,nofootinbib,superscriptaddress]{revtex4-1}
\usepackage{amsmath,graphicx,latexsym,aeguill}
\usepackage{xcolor}
\usepackage{physics}
\usepackage[colorlinks=false,bookmarks=false,citecolor=blue,linkcolor=blue,urlcolor=blue, hidelinks=true]{hyperref}
\begin{document}

\title{Thermal and magnetic field stability of holmium single atom magnets}
\author{Fabian Donat Natterer}
\affiliation{Institute of Physics, {\'E}cole Polytechnique F{\'e}d{\'e}rale de Lausanne, Station 3, CH-1015 Lausanne}
\author{Fabio Donati}
\affiliation{Center for Quantum Nanoscience, Institute for Basic Science, Seoul 03760, Republic of Korea}
\affiliation{Institute of Physics, {\'E}cole Polytechnique F{\'e}d{\'e}rale de Lausanne, Station 3, CH-1015 Lausanne}
\affiliation{Department of Physics, Ewha Womans University, Seoul 03760, Republic of Korea}
\author{Fran\c{c}ois Patthey}
\affiliation{Institute of Physics, {\'E}cole Polytechnique F{\'e}d{\'e}rale de Lausanne, Station 3, CH-1015 Lausanne}
\author{Harald Brune}
\affiliation{Institute of Physics, {\'E}cole Polytechnique F{\'e}d{\'e}rale de Lausanne, Station 3, CH-1015 Lausanne}
\begin{abstract}
We use spin-polarized scanning tunneling microscopy to demonstrate that Ho atoms on magnesium oxide exhibit a coercive field of more than 8~T and magnetic bistability for many minutes, both at 35~K. The first spontaneous magnetization reversal events are recorded at 45~K for which the metastable state relaxes in an external field of 8~T. The transverse magnetic anisotropy energy is estimated from magnetic field and bias voltage dependent switching rates at 4.3~K. Our measurements constrain the possible ground state of Ho single atom magnets to either $J_z = 7$ or 8, both compatible with magnetic bistability at fields larger than 10~mT.
\end{abstract}
\maketitle

Individual Holmium atoms on magnesium oxide (MgO) thin films were identified as the first example of stable single atom magnets~\cite{don16}. They exhibit long-lived magnetic quantum states giving rise to remanent magnetization, representing the ultimate size limit of a magnetic bit~\cite{don16}. While the reading and writing of individual Ho atoms was demonstrated~\cite{nat17}, the thermal stability of their magnetization, their coercitive field, and their magnetic ground state remain open issues. A further single atom magnet system, Dy on graphene, was discovered~\cite{bal16}. It exhibits steps in magnetization curves due to magnetic level crossing, also characterizing molecular magnets. The system Ho/MgO, however, is exceptional as there is no evidence of level crossing up to 8~T. Consequently, it can have a much larger remanence and an extremely high coercitive field.

The first two single atom magnet systems were explored using X-ray magnetic circular dichroism (XMCD)~\cite{don16, bal16}. However, systems with slow magnetic relaxation times are potentially perturbed by photon-induced secondary electrons that activate additional magnetic relaxation channels. The intrinsic relaxation time may be derived from flux dependent measurements and extrapolation to zero photon flux~\cite{dre14}. Independently of the method, measurements on ensembles with very long lived magnetic quantum states may probe state populations that are kinetically limited and thus deviate from thermodynamic equilibrium. As our measurements in the present letter show, the magnetic saturation of an ensemble of Ho atoms on MgO might indeed be very difficult to achieve in a field sweep. The magnetization curves recorded by XMCD may consequently show a lower saturation, smaller hysteresis, and therewith derived magnetic moments would correspondingly appear smaller. This may explain the discrepancy to the Ho moment measured by electron-spin resonance~\cite{nat17}.

Scanning tunneling microscopy (STM) has been used to probe the magnetic state of Ho atoms via tunnel magneto resistance (TMR) at bias voltages below a threshold of 73~meV~\cite{nat17}. Above this threshold the magnetic quantum states of Ho can be written. Here, we use this technique to establish new boundaries for the temperature and field dependent magnetic stability and for the magnetic ground state of Ho/MgO.

\begin{figure}
\includegraphics[width = 8.6 cm]{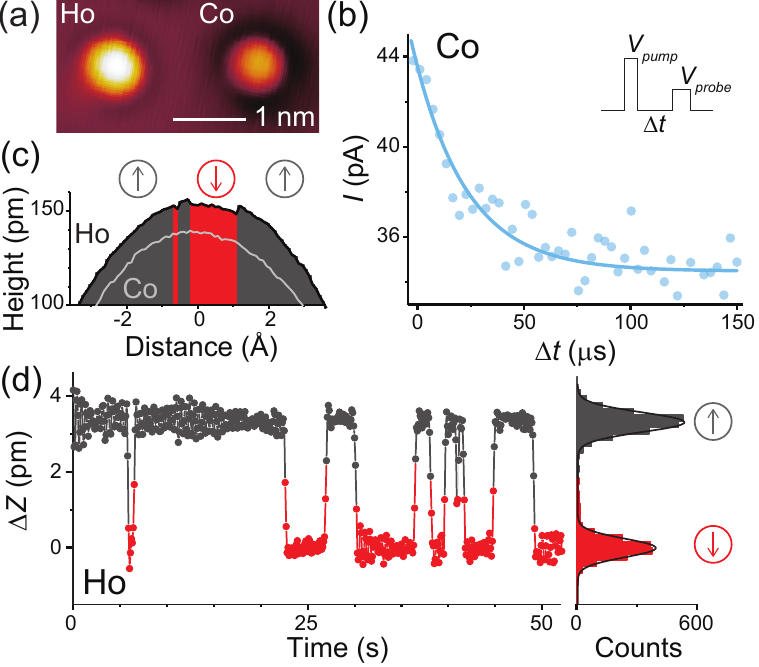}
\caption{(a) STM image of co-adsorbed Ho and Co atoms on MgO/Ag(100). (b) Pump-probe measurement of the Co atom demonstrating spin-polarization (SP) of the tip along the surface normal. (c) Line profile of a Ho atom measured with a SP-tip at 150~mV showing apparent height contrast between the two magnetic states, in contrast to the profile across the paramagnetic Co atom. (d) Telegraph noise in apparent height due to magnetic switching of a Ho atom. The histogram indicates a preference for one magnetization direction, identifying the magnetic ground state being UP in the external field pointing UP ((a) -- (d) $B_z = 4$~T, $T_{\rm STM} = 4.3$~K, $V = 150$~mV, $I = 50$~pA, (b) $V_{\rm pump} = 150$~mV, $V_{\rm probe} = 40$~mV).}
\label{STM}
\end{figure}

\begin{figure*}
\includegraphics[width = 17 cm]{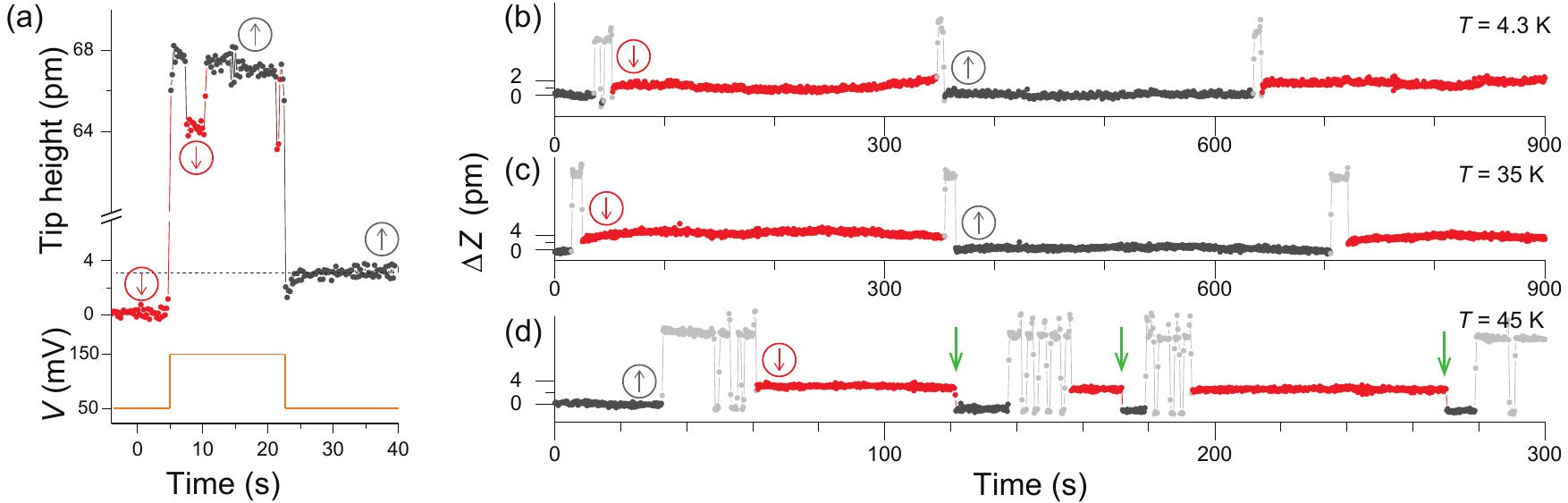}
\caption{Magnetic bistability of Ho atoms at elevated temperature and in an external field of $B_z = 8$~T pointing UP. (a) Intentional switching by increasing the voltage to $V = 150$~mV. (b) Tip-height time trace recorded at 4.3~K. The Ho moment is stable in both orientations. (c) -- (d) Same as in (b) but measured at 35 and 45~K, respectively. While the metastable state remains essentially unaffected at 35~K, the onset of spontaneous reversals (green arrows) is seen at 45~K. (b) -- (d) monitoring bias $V \le 50$~mV, switching bias $V \ge 120$~mV. A smoothing filter results in an effective bandwidth of 3~Hz.}
\label{temp}
\end{figure*}

The measurements were made with a homemade STM operating at temperatures from 0.4~K to 50~K and in an out-of-plane magnetic field $(B_z \parallel z)$ of up to 8~T~\cite{cla05, sup}. The MgO layer was grown on Ag(100) following recent literature~\cite{pal14, sup}. We spin-polarize (SP) a W-tip by transferring surface-adsorbed cobalt atoms to the tip apex (see Fig.~\ref{STM}(a)). The Co atoms are distinguished from Ho by their apparent height and 58~meV spin-excitation signature~\cite{rau14}. We test the tip's spin-polarization along the $z$-axis via pump-probe spectroscopy~\cite{lot10} on Co that has a very large out-of-plane anisotropy on MgO~\cite{rau14}. Figure~\ref{STM}(b) shows the typically achieved spin-polarization of 26~\% on Co and a spin-relaxation time of $(25 \pm 4)$~${\rm \mu}$s at 4~T, in agreement with earlier results~\cite{bau15}. Once the tip's spin-sensitivity is established, we use it to read the magnetic state of Ho single atom magnets via the TMR~\cite{nat17}. We measure at constant current, and therefore a change in the magnetic state of the Ho atom leads to a sudden jump in tip-height (see Fig.~\ref{STM}(c)).

Figure~\ref{temp} displays time traces of the tip-height above Ho monitoring its magnetic state, UP or DOWN, at different temperatures. In our measurements the external field of 8~T is pointing UP (out-of-plane), therefore the UP state is the ground and DOWN the metastable state. Panel (a) illustrates the preparation of a desired quantum state. We start with the magnetization pointing DOWN. In order to reverse the magnetization to UP, we increase the tunnel voltage to 150~mV, leading to several consecutive switches between the two states and we lower the tunnel voltage back to 50~mV, once the atom is in the desired UP state. The probe voltage is lower than any of the electron-activated switching thresholds~\cite{nat17} and yields a tip-height contrast of typically $2-4$~pm between the Ho states. Figure~\ref{temp}(b) shows a tip-height time trace measured at 4.3~K. The black and red colored segments correspond to the Ho atom in the UP and DOWN state, respectively. The continuous tip-height trace shows that either state is stable for the several hundreds of seconds observation time. This is remarkable, given the large Zeeman energy difference of 9.36~meV between UP and DOWN state, estimated from the Ho magnetic moment of $(10.1 \pm 0.1)~{\rm \mu}_{\rm B}$~\cite{nat17}. The metastable Ho quantum state withstands the opposing external magnetic field, which translates into an exceptional coercivity of at least $8$~T for this single atom magnet.

Even more striking, magnetic bistability also extends to significantly higher temperatures. Figure~\ref{temp}(c) shows the tip-height time trace at 35~K. As for the case at 4.3~K, either state, UP and DOWN, is stable. We only see deliberate switching induced by current pulses at $V \ge 120$~mV. Figure~\ref{temp}(d) finally shows the onset of a few spontaneous magnetization reversals (green arrows) from the metastable to the ground state at around 45~K. We measure a mean lifetime of the metastable state of $(66 \pm 33)$~s. The high temperature required to activate thermal-induced switching further confirms the weak interaction between the magnetic quantum states and lattice vibrations. Note that above 50~K, Ho atoms start moving away from the oxygen to the bridge site of the MgO lattice~\cite{fer17}, where they lose their single atom magnet properties, thus defining a natural temperature limit above which the system is irreversibly transformed, in agreement with previous observations~\cite{don16}.

\begin{figure*}
\includegraphics[width = 15 cm]{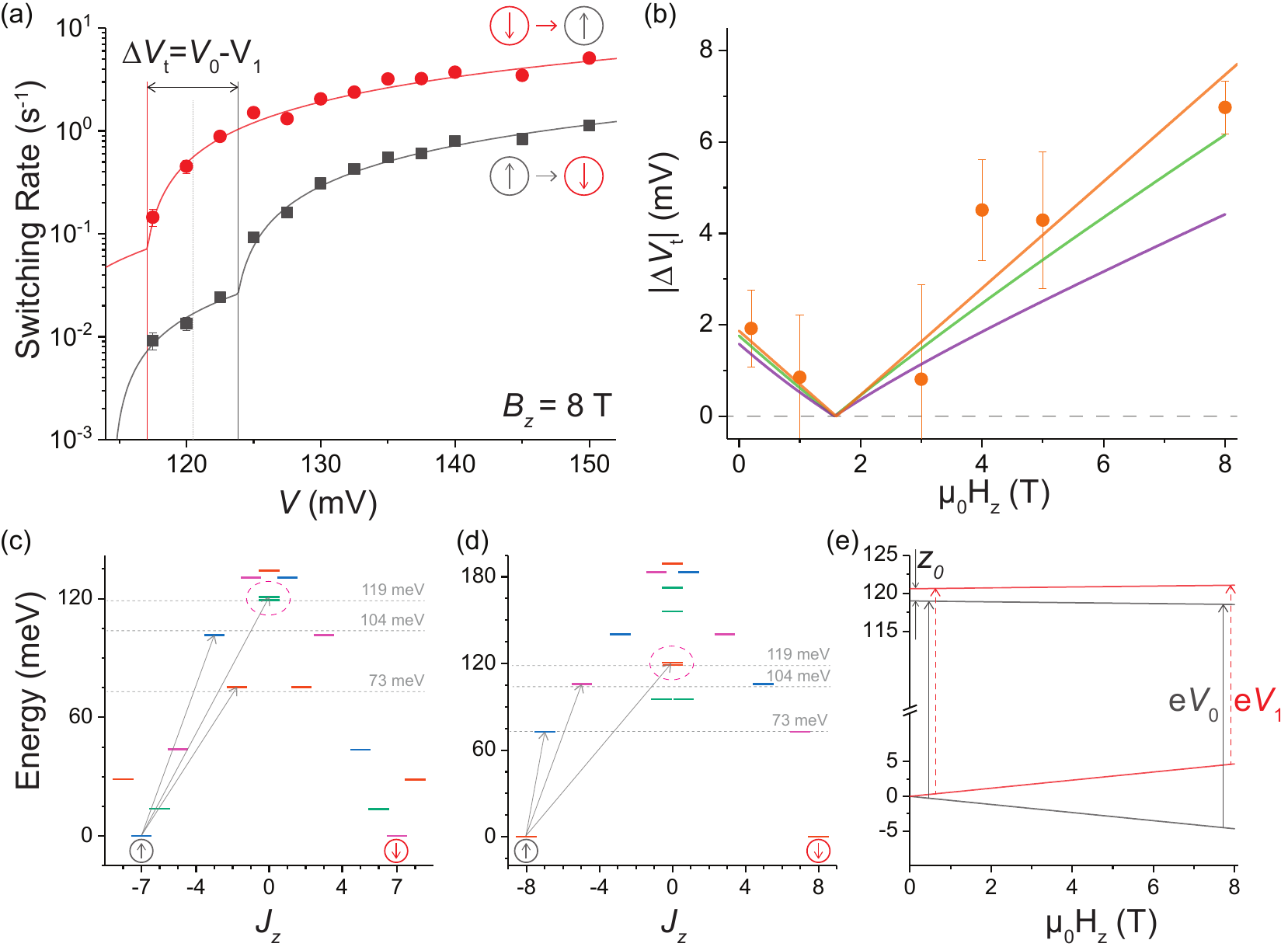}
\caption{Field dependent magnetic switching. (a) State resolved switching rate at $B_z = 8$~T, measured at $1.5$~nA. The switching rates of the UP and DOWN states show the onset of a new rate increasing threshold at two different bias voltages $V_0$ and $V_1$. The solid lines represent a piecewise linear fit~\cite{sup}. (b) Magnetic field dependence of the threshold difference, $\abs{\Delta V_{\rm t}}$, indicating a zero-field offset $z_0$. The lines represent a fit to $\abs{\Delta V_{\rm t}}= \abs{z_0 - 2\Delta m B_z}$ (orange), show model calculations of a $J_z =7$ (green) and $J_z =8$ (purple) ground state. The kink occurs when $V_1>V_0$. Energy level diagram for (c) $J_z =7$ and (d) $J_z =8$ ground state at $10$~mT~\cite{sup}. The arrows and the horizontal line denote the three thresholds reported in Ref.~\cite{nat17}. The dotted circles highlight the intermediate state with zero-field-splitting $z_0$. Levels of equal color are connected via the fourfold symmetry. (e) Zeeman energy sketch of the ground and intermediate states in dependence of applied field. Arrows indicate transitions from the ground (black) or the excited state (red) to the zero-field split intermediate state. All measurements were recorded at $4.3$~K.}
\label{sw}
\end{figure*}

In order to illuminate possible origins for the extraordinary stability of the Ho moment, we investigate the magnetic field dependent switching behavior for bias voltages above the switching threshold. We determined the switching rate from tip-height time traces, such as the one shown in Figure~\ref{STM}(d), and separately quantify the lifetime of either state. Figure~\ref{sw}(a) shows the state resolved switching rate, from UP to DOWN and vice versa, at our maximal field of $B_z = 8$~T~\cite{sup}, measured for bias voltages between 117 and 150~mV (at 1500~pA). The switching rates of the ground (black) and metastable (red) state grow monotonically with applied bias voltage due to the increasing fraction of tunnel electrons exceeding this switching threshold. We observe the onset of a rate increasing threshold at a bias voltage $V_0 = 124$~mV for the ground, and at $V_1 = 117$~mV for the metastable state. The average value coincides well with the third threshold of $119 \pm 1$~mV reported in Ref.~\cite{nat17}. The overall reduced switching rate at higher fields suggests an effective suppression of quantum tunneling of magnetization (QTM), possibly due to the improved purity of states~\cite{ung16}. The threshold difference, $\abs{\Delta V_{\rm t}} = \abs{ V_0 - V_1}$ (see Fig.~\ref{sw}(e)), between the switching of the ground and metastable state, first shrinks from a distinct offset to zero and then increases with magnetic field when $V_1 > V_0$. We thus assume a zero-field splitting for the intermediate state and a linear Zeeman trend to which we fit the threshold difference $\abs{\Delta V_{\rm t}} = \abs{z_0 - 2\Delta m B_z}$ (orange line in Figure~\ref{sw}(b)). We find a zero-field offset of $z_0 = (1.8 \pm 0.8)$~mV and a magnetic moment difference, $\Delta m$, between the ground and intermediate state of $(9.5 \pm 1.1)$~$\mu_{\rm B}$. The 1.8~meV zero-field splitting in the intermediate state doublet results from the transverse anisotropy induced by the $C_{4v}$ symmetry of the oxygen adsorption site. The measured $\Delta m$ is compatible with the Ho ground state moment found previously~\cite{nat17}. Earlier XMCD measurements which derived $J_z = 4.7$ seemingly underestimated the magnetic moment of Ho~\cite{don16}, possibly due to non-equilibrium states probed at 8~T.

In view of these observations, we revise the Ho level scheme~\cite{don16} and propose two possible models, with distinct ground state doublet $J_z = 7$ or 8~\cite{sup}. The $J_z = \pm 7$ case is symmetry protected from QTM and naturally compatible with magnetic bistability on the $C_{\rm 4v}$ adsorption site~\cite{hub14, don16}. The corresponding magnetic moment $m_{\rm 4f} = g J_z\mu_{\rm B} \approx 8.8$~$\mu_{\rm B}$, however, requires the presence of additional 1.3~$\mu_{\rm B}$ in an outer 6{\it s}/5{\it d} shell to balance the measured Ho moment. We contrast the latter model to a $J_z = \pm 8$ scenario, for which the magnetic moment would be mostly localized on the 4{\it f} orbitals~\cite{nat17}. The latter ground state is stable only at finite magnetic field. We reject cases of lower $J_z$ for which the total magnetic moment observed by ESR-STM cannot be matched.

In the effective Spin-Hamiltonian analysis, detailed in Ref.~\cite{sup}, we formulate constraints for the magnetic level structure that are compatible with our $\abs{\Delta V_{\rm t}}$ data and earlier observations. We first tune the Stevens parameters to match the three previously reported switching thresholds~\cite{nat17}, while respecting the absence of level crossings within the ground state doublets up to $9$~T~\cite{don16}. The two corresponding energy level diagrams are shown for 10~mT in Figs.~\ref{sw}(c) $(J_z = 7)$ and \ref{sw}(d) $(J_z = 8)$. We additionally calculate the probability of spin reversal via transitions through intermediate states~\cite{gau13, sup}. Within these models, the state switching described in this work would correspond to the transition between the legacy states $J_z = \pm7 \rightarrow \pm2$ or $J_z = \pm8 \rightarrow \pm4$ of an ideal uniaxial anisotropy case. In agreement with the present observations, both models have a large probability for reversing the Ho spin through the third threshold for all field values, whereas the lowest two transitions become strongly suppressed in an applied magnetic field. Within the error bar, both models reproduce our $\abs{\Delta V_{\rm t}}$ data. The $J_z = 7$ case is naturally bistable through symmetry protection, whereas the $J_z= 8$ state relies on the large $J$ quantum number and strong uniaxial anisotropy protecting the ground states~\cite{ses93}. The Zeeman energy overwhelms the small transverse anisotropy term (Stevens parameter $B_4^4 <4$~$\mu$eV~\cite{sup}) for large $J_z$ values already at a few mT, leading to nearly degenerate ground states but to tunnel-split intermediate states (low $J_z$). The externally applied field is thus able to lift the mixing of the ground state doublet and effectively suppress QTM. Holmium atoms of this $J_z = 8$ scenario would in fact not show bistability but rather be in a superposition of UP and DOWN state at zero-field. In this case, the remanence observed in XMCD magnetization loops would have resulted from the slow dynamic of QTM-related switching that follows from the small transverse anisotropy. The weak coupling to MgO phonons may also suppress QTM, as observed for molecular magnets coupled to nanomechanical resonators~\cite{gan13}. In addition to the reorientation of the magnetic moment via high energy electrons from the STM tip and via thermal excitations, the state mixing may be a further alternative of how to control the spin-state via the QTM~\cite{gat03} for an otherwise proven stable and coercive single atom magnet.

The combined thermal and magnetic field stability reported here for the magnetic quantum states of single Ho atoms on MgO(100) surpasses the presently most stable molecular magnet dysprosocenium, reaching 60~K, but with very little remanence of < 1~\% $M_{\rm  sat}$ and short magnetic lifetime at high magnetic fields~\cite{goo17}. The suggested magnetic ground states for Ho are both intriguing candidates for quantum information processing and high density data storage applications; the $J_z = \pm 7$ state is ideally protected, while the $J_z = \pm 8$ is stable only at finite field, whereas at zero field it exhibits superposition states that are easy to manipulate.

\begin{acknowledgements}
We thank A. Singha and S. Rusponi for discussions and our mechanical and electronic workshop for expert assistance. F.D.N. acknowledges support from the Swiss National Science Foundation under project number ${\rm PZ00P2\_167965}$.
\end{acknowledgements}

\bibliographystyle{apsrev4-1}
\bibliography{ms_14}

\widetext
\clearpage
\appendix
\setcounter{figure}{0}
\renewcommand{\thepage}{S\arabic{page}}
\renewcommand{\thesection}{S\arabic{section}}
\renewcommand{\thetable}{S\arabic{table}}
\renewcommand{\thefigure}{S\arabic{figure}}

\section{Supplemental Information}
\subsection{Sample Preparation}
We performed the experiments with our home-built low-temperature STM~\cite{cla05} operating at temperatures ranging from $440$~mK to 50~K and in magnetic fields $B_z$ of up to 8~T applied along the surface normal $z$. For temperatures exceeding 12~K, we substitute the helium of the cryostat in the insert with solidified nitrogen. The weak thermal link between the $^3$He pot and the insert is regularly used for milli Kelvin operation but it also enables a slow warm-up to about 50~K taking 3-4 days, while the outer cryostat holding the magnets and the STM shields is kept at liquid helium temperature. The magnesium oxide (MgO) film was grown on atomically clean Ag(100) by exposing the sample held at 770~K to an Mg flux from a Knudsen cell in an oxygen partial pressure of $10^{-6}$~mbar and at a growth rate of about 0.5~monolayers per minute, similar to Ref.~\cite{pal14}. After transferring the sample into the cooled STM, we simultaneously dosed cobalt and holmium from high purity rods using an electron-beam evaporator onto a sample held at 3~K. Under these conditions, the atoms come to rest at or very close to their impact site where they remain immobile.

\subsection{Ho single atom switching and sources of error}
\begin{figure*}[h]
\includegraphics[width = 17 cm]{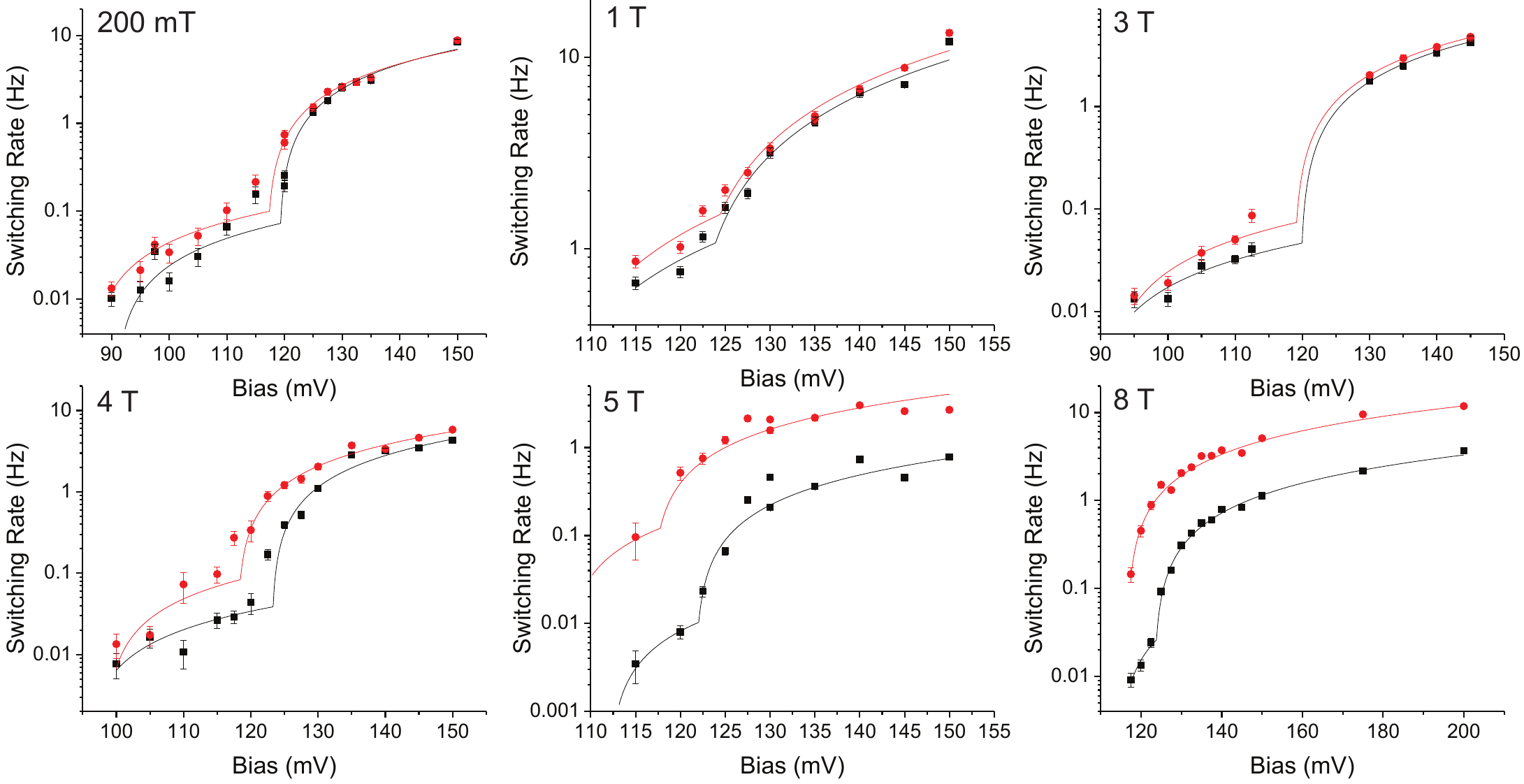}
\caption{Field dependent magnetic switching in the bias voltage range from $90 - 200$~mV measured at 1500~pA, except for the data set at 3~T, which was measured at 500~pA. The rate increasing threshold at 119~mV is visible for all data sets. The solid lines are piecewise linear fits ($T_{\rm STM} = 4.3$~K).}
\label{SI_field}
\end{figure*}

In view of the long magnetic lifetime and for practical reasons, we monitor the Ho states via changes in the tip-height at constant current. This allows us to continuously track the Ho atom even in presence of thermal drifts, for instance at the elevated temperatures shown in Fig.~\ref{temp} of the main text. We determined the switching rate from the lifetime of the Ho atom in the UP and DOWN state separately. An exponential fit to the residence time distribution in each state yields the state resolved lifetime {\it viz} the switching rate at a specific bias voltage, shown in Fig.~\ref{sw}(a) of the main text and in Fig.~\ref{SI_field}. We fit the magnetic field dependent switching rates for the UP and DOWN state to a piecewise linear trend of form $R = \sum{a_i \tfrac{V - V_i}{V_i}}$, where $V_i$ is the threshold voltage and $a_i$ a slope coefficient. Figure~\ref{SI_field} shows the complete data set from which the threshold differences, $\abs{\Delta V_{\rm t}}$, are reported in Figure~\ref{sw}(b) of the main text. The voltage difference was chosen to report values which are independent of eventual voltage offsets.

The main sources for error in our analysis stem from the small number of switching events at low bias voltages, and from the about 10~pm peak-to-peak $z$-noise level, which is more critical at higher switching rates due to the required averaging times of 20 to 100~ms in view of the oftentimes only 2~pm tip-height contrast. We accounted for missed events via $R_{\rm true} = R_{\rm meas}/(1+R_{\rm meas}*\tau)$, where $\tau$ is the effective dead-time (maximum of preamplifier bandwidth and averaging time), and $R_{\rm meas}$ is the measured switching rate~\cite{mul73}.

The vertical error bars in the Zeeman plot of Fig.~\ref{sw}(b) represent the standard deviation of the respective lifetime fits. Note also that owing to the actual orientation of the tip moment of individual micro-tips, an inverted tip-height signature for the UP and DOWN state may be possible (see for instance Figs.~\ref{temp}(b) -- (d) of the main text).

\subsection{Spin Hamiltonian analysis}
We model the magnetic level structure of Ho atoms using a reduced Spin Hamiltonian of the form:

\begin{equation}
H = H_{\rm cf} + H_{\rm Z},
\end{equation}

which includes the effects of crystal field and of the external magnetic field, respectively. We use this Spin Hamiltonian to describe the splitting of the lowest multiplet of magnetic states. For the Ho atom in the $4f^{10}$ configuration~\cite{don16}, the lowest multiplet consists of states with a total magnetic moment $J = 8$. This quantum number determines the corresponding multiplicity of the states $2J + 1 = 17$. For the Ho atoms adsorbed on the four-fold symmetric O site~\cite{fer17}, only the crystal field operators allowed by the $C_{4v}$~symmetry are included:

\begin{equation}
H_{\rm cf} = B^{2}_{0}\, \hat{O}^2_0\, +  B^{4}_{0}\, \hat{O}^4_0\, +  B^{4}_{4}\, \hat{O}^4_4\, +  B^{6}_{0}\, \hat{O}^6_0\, +  B^{6}_{4}\, \hat{O}^6_4\, .
\end{equation}

The value of the $B^{n}_{m}$ coefficients determines the zero-field splitting of the magnetic states. The uniaxial operators ${\hat O}^{n}_{0}$ commute with the $z$--projected total moment operator $\hat{J}_z$, therefore they preserve the corresponding quantum number $J_z$. Conversely, the transverse operators ${\hat O}^{n}_{4}$ mix states separated by $\Delta J_z = 4$. Consequently, all the conjugated states with even value of $J_z$ are coupled into pairs of singlets with vanishing expected value of $\langle J_z \rangle$.

We evaluate two alternative models with respect to the one presented in Ref.~\cite{don16}, which was obtained from multiplet calculations. As described in the text, the values of the $B^{n}_{m}$ coefficients have been chosen to constrain the ground state to either $J_z = 7$ or 8, as states with lower $J_z$ cannot account for the value of the total magnetic moment $m_{\rm Ho} = (10.1 \pm 0.1)$~$\mu_{\rm B}$ found in former STM experiments~\cite{nat17}.

The Zeeman term in the Hamiltonian describes the interaction between the Ho spins and the external magnetic field. It reads:

\begin{equation}
H_{\rm Z} = -\overrightarrow{m}_{\rm Ho}\cdot \overrightarrow{B}=-\left(\overrightarrow{m}_{4f}+\overrightarrow{m}_{5d6s}\right)\cdot \overrightarrow{B} =- g_{\rm eff}\overrightarrow{J}\cdot \overrightarrow{B} \mu_{\rm B},
\end{equation}

where we explicitly decompose the total magnetic moment into the contribution of $4f$ and $6s5d$ electrons. We assume the spin moments in the outer orbitals to be much smaller than and in parallel coupling with the $4f$ electron spin. Within this assumption, a non vanishing $m_{5d6s}$ does not affect the multiplicity of the magnetic levels and its interaction with the external field can be accounted for in the effective value of the Land\'{e} factor $g_{\rm eff}$. In the present experiments the external field is always oriented perpendicular to the surface, hence $\overrightarrow{B} = B_z \overrightarrow{e_z}$:

\begin{equation}
H_{\rm Z}=-g_{eff} J_z B_z \mu_{\rm B}.
\end{equation}

For a ground state with $J_z = 8$, the magnetic moment of the $4f$ shells $m_{4f} = g_{4f} J_z = 10$~$\mu_{\rm B}$ essentially suffices to obtain the required value of $m_{\rm Ho} = (10.1 \pm 0.1)$~$\mu_{\rm B}$. In this case, $m_{5d6s}$ is negligible and $g_{\rm eff} = g_{4f} = 1.25$. Conversely, a ground state with $J_z = 7$ has $m_{4f}=g_{4f} J_z = 8.8$~$\mu_{\rm B}$, therefore a finite $m_{5d6s} = 1.3$~$\mu_{\rm B}$ is required to match the total magnetic moment of the Ho atom. We include this moment assuming a $g_{\rm eff} = m_{\rm Ho}/7 = 1.44$.

We further constrain the hierarchy of the magnetic levels using the energy thresholds for magnetization switching obtained from previous~\cite{nat17} and present STM experiments. These thresholds can be evaluated in the Spin Hamiltonian model by calculating the probability of inducing a reversal via spin--excitation events. The reversal path involves an excitation towards an intermediate state with a single electron scattering process, and a successive decay to a spin state with opposite sign. The probability of a reversal path that starts from an initial state $J_{z, i}^\uparrow$ through an intermediate state $J_{z, j}$ ending in any final state $J_{z, f}^\downarrow$ with energy $E_f < E_j$ is calculated as the product of the two spin-excitation events:

\begin{equation}
\mathfrak{P}_{\rm rev}(j)=\mathfrak{P}(J_{z,i}^{\uparrow}\rightarrow J_{z,j})\times\sum_{f\in \{\downarrow, E_f<E_j\}}\mathfrak{P}(J_{z,j}\rightarrow J_{z,f}^\downarrow)\, ,
\end{equation}

where the sum includes all the possible final states with inverted magnetization, and the spin excitation probability is calculated as described in Ref.~\cite{hir07}. We assume that, from any of these states, the system further relaxes to the lowest long living $J_z^\downarrow$ state. The reversal probability is then evaluated for all the intermediate $J_{z, j}$ states as a function of magnetic field.

\begin{figure*}
\includegraphics[width = 16 cm]{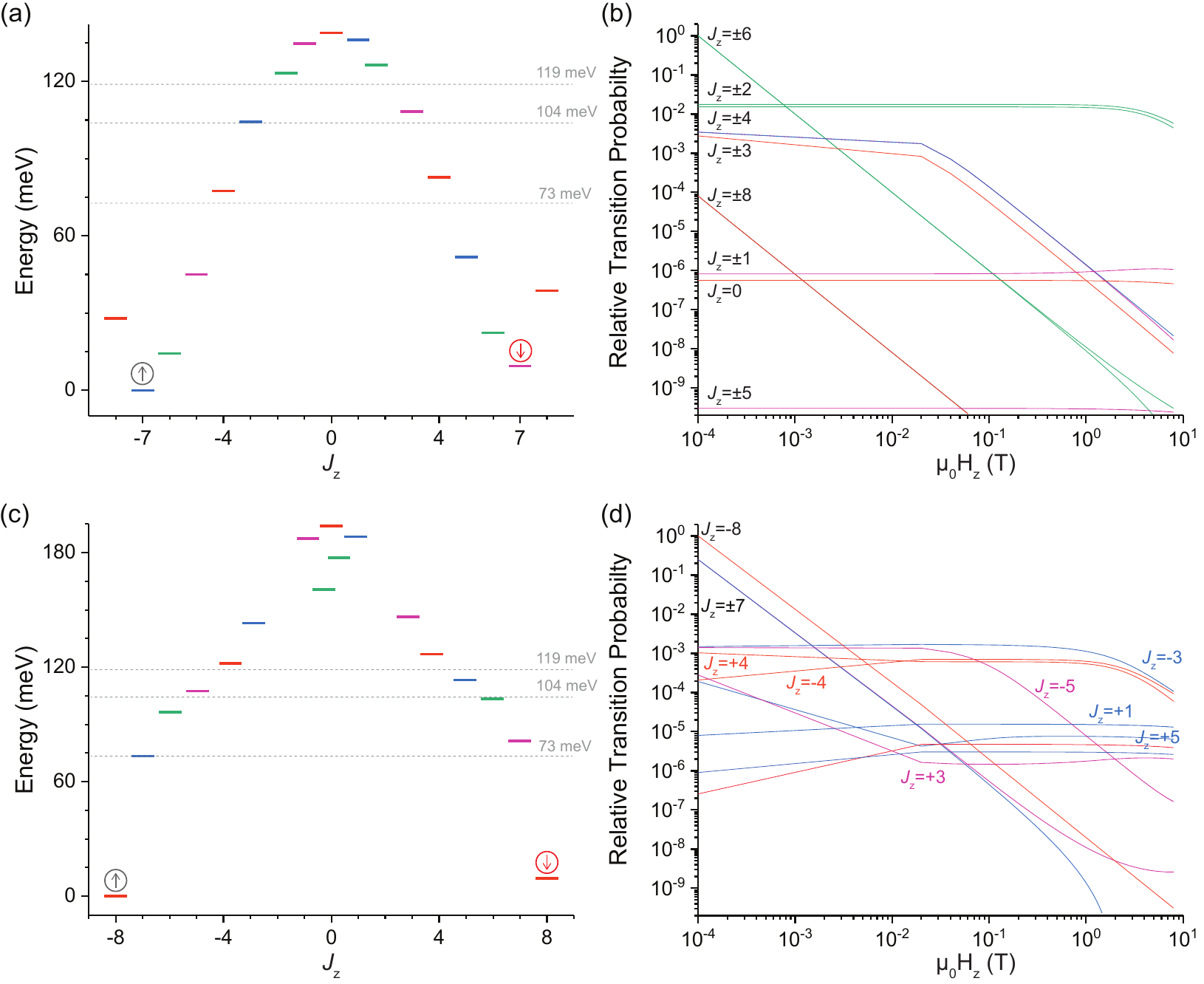}
\caption{Level diagram comparison for Ho/MgO in the (a) $J_z = 7$ and (c) $J_z = 8$ ground state at a field of 8~T. Transition probabilities for (b) $J_z = 7$ and (d) $J_z = 8$ via the labeled intermediate states, showing the effective suppression of reversal pathways at increasing magnetic field.}
\label{SI_probability}
\end{figure*}

The knowledge of the probability for spin reversal allows us to fine tune the energy of the three intermediate states that provide the most efficient reversal path. Therefore, we varied the crystal field parameters for the two models until the energy of these states match the thresholds identified in \cite{nat17}. In addition, we require the intermediate level at $119$~meV to be a tunnel-split pair of singlet states separated by an energy $z_0 = 2$~meV. These states allow the transition path explored in the present experiment and a zero-field energy separation is required to match the results described in Fig.~\ref{sw}(b). Table~\ref{CF} summarizes the coefficients of the Stevens operators generating the level structures depicted in Figs.~\ref{sw}(c) and (d) and Figs.~\ref{SI_probability}(a) and (c).

\begin{table}[h]
\caption{Crystal field parameters used in Spin Hamiltonian calculations for the two models described in the text.}
\begin{center}
\begin{tabular}{|c|c|c|c|c|c|}
\hline
& $B^2_0$ & $B^4_0$ & $B^4_4$ & $B^6_0$ & $B^6_4$ \\
\hline
$J_z =7$ & $-670~\,\mu$eV & $+1.6~\,\mu$eV & $+250$~neV & $+4$~\,neV & $-1.5$~\,neV \\
\hline
$J_z =8$ & $-835~\,\mu$eV & $-100$~neV & $+3.7~\,\mu$eV & $+8.6$~\,neV & $0$~\,neV \\
\hline
\end{tabular}
\end{center}
\label{CF}
\end{table}

These figures show the level splitting at $B_z=10$~mT and $8$~T, respectively. For the model with $J_z = 7$ ground state, the strongest transition corresponds to the excitation to the coupled states with $J_z = \pm2$. In the $J_z = 8$ model, the reversal has to occur via $J_z = \pm 4$ states given that the other split doublets have vanishing matrix elements for a spin transition. Constraining a $z_0=2$~meV for these states requires a larger transverse term than for the case of $J_z = 7$. In contrast to the $J_z = 7$ model, the transverse term of the $J_z = 8$ scenario may indeed lead to a non-remanent tunnel split ground state at low fields. However, for the value adopted in this model, the reversal probability is sufficiently quenched (namely lower than that obtained from other relaxation pathways) already at $B_z = 10$~mT, thus in agreement with the long magnetic lifetime found in XMCD measurements~\cite{don16}. The evolution of the transition probability as a function of magnetic field is shown in Figs.~\ref{SI_probability}(b) and (d). It illustrates the action of the magnetic field in suppressing reversal pathways via the respective intermediate states.
\end{document}